\documentclass[aps,pra, twocolumn]{revtex4}

\usepackage{graphicx}
\usepackage{amssymb}
\usepackage{amsmath}
\usepackage{colordvi}
\usepackage{color}

\newcommand{\cC}{{\cal C}}

\newcommand{\cE}{{\cal E}}
\newcommand{\cH}{{\cal H}}

\newcommand{\tmu}{\tilde{\mu}}
\newcommand{\tn}{\tilde{n}}

\newcommand{\RE}{\textrm{Re}}

\newcommand{\vep}{\varepsilon}
\newcommand{\rev}[1]{\textcolor{black}{#1}}
\newcommand{\sech}{\, \mathrm{sech}}

\begin{document}

\title{Double-flattop  quantum droplets in \rev{low-dimensional} Bose-Bose mixtures}

\author{Yaroslav V. Kartashov$^1$ and Dmitry A. Zezyulin$^2$}

\affiliation{$^1$Institute of Spectroscopy, Russian Academy of Sciences, Troitsk, Moscow, 108840, Russia\\\medskip
$^2$School of Physics and Engineering, ITMO University, St. Petersburg  197101, Russia}


\date{\today}

\begin{abstract}

We predict the existence of double-flattop quantum droplets in atomic Bose-Bose mixtures. Solutions of this type have two flattop regions of nearly uniform atomic density corresponding to a compressed central core  surrounded by a rarefied layer. The birth of these double-flattop quantum droplets is analytically described using a perturbation theory, which in the leading order reduces the problem to the cubic nonlinear Schr\"odinger equation. Its properties are then used to predict the shape of double-flattop solutions and draw the conclusions about their stability. The analytical results apply to one- and multidimensional quantum droplets, provided that the energy density satisfies certain conditions. Using the numerical continuation from the asymptotic limit, we obtain the families of one- and two-dimensional double flattop quantum droplets and confirm the stability of the nodeless states of this type.

\end{abstract}

\maketitle

\section{Introduction}

In the lowest-order approximation of Bogoliubov theory for a dilute gas of weakly interacting Bose particles \cite{Bogo1947}, the ground-state energy density is proportional to the squared particle density, and the coefficient of proportionality is positive or negative for repulsive or attractive interactions, respectively.  For a nonuniform Bose-Einstein condensate (BEC), this approximation can be developed into the well-known Gross-Pitaevskii equation \cite{Pita}, which is widely and  successfully used for theoretical description of BECs in a variety of realistic settings. At the same time, it has been predicted that in dilute two-component Bose-Bose mixtures, quantum corrections to the ground state energy  \cite{LHY} may become non-perturbative and, moreover, may enter into competition with the meanfield cubic nonlinearity \cite{Petrov2015}. A delicate balance between meanfield and beyond-meanfield nonlinearities enables the formation of quantum droplets, i.e., self-bound liquidlike states that can exist in stable form even in free space  \rev{\cite{Bulgac,Michi2002}}. Quantum corrections play in this case a constructive role by stabilizing the condensate against collapse predicted by the   meanfield theory. Another prominent property of quantum droplets (which further emphasizes  the analogy with usual liquids) is the incompressibility: irrespective of the number of condensed atoms, the particle density cannot exceed a certain maximal value. As a result, for large numbers of  atoms, quantum droplets tend to develop the so-called flattop shapes, i.e.,  broad plateaus of nearly uniform density \rev{(as it happens also in optical materials with competing nonlinearities \cite{Quiroga})}. Soon after the theoretical prediction, quantum droplets in Bose-Bose mixtures were observed in several experiments \cite{Cabrera2018, Semeghini2018, Cheiney2018, 39K, collisions, hetero}. Analogous quantum liquidlike states, resulting from the balance between repulsive and attractive terms, have been also observed in single-component dipolar condensates \cite{Schmitt2016, FerrierBarbut2016, Baillie2018}. Broad reviews of the progress made in rapidly developing area of multidimensional quantum droplets can be found in \cite{Luo2021, Malomed2021, Khan2022, Multidim2019, MaloBook, Bottcher2021}.

The beyond-meanfield theory of Bose-Bose mixtures can be generalized to low-dimensional liquids    \cite{Petrov2016, Parisi1, Parisi2}, where the functional form of the energy density differs from that in the three-dimensional (3D) case, that results in different modifications of the Gross-Pitaevskii equation with functional form of nonlinearity  depending on dimensionality of the problem. Steady-state solutions corresponding to quantum droplets emerging in such models have been in the focus of numerous theoretical studies. The simplest solutionscan be obtained by assuming that density distributions in both components are proportional  \cite{Li2018, Petrov2016, AstaMalo, Katsimiga2023, Katsimiga2023CM, Edmonds2023, Khan2022, Abdullaev, Otajonov, Paredes, Dong2024,DongFan2024, Kartashov2019, Kartashov2018, Kartashov2021, Gangwar2024,Bougas2024,Flynn2023}. This assumption substantially simplifies the problem but, at the same time, it drastically restricts the set of available quantum droplet solutions. More complex, essentially two-component states, which cannot be described in terms of  a single  wavefunction, have also been addressed in several recent studies \cite{Mithun2020, Kartashov2020, Mista2021, Kartashov2022, KartashovZ2024,Gangwararx,Englezos2024,Charalampidis2024}, but in general they received insufficient attention, especially in multidimensional geometries, where exploration of new types of self-sustained states that can be dynamically stable, is a task of considerable interest. Particularly interesting solutions, which cannot be implemented in the scalar case, correspond to structured quantum droplets with essentially different density distributions in two components. For instance, two wavefunctions describing components of the quantum droplet may have different parities \cite{KartashovZ2024} (in the 1D case) or different topological charges \cite{Kartashov2020} (for vortex states in the 2D case).

The goal of this paper is to introduce a previously unexplored class of \textit{double-flattop} quantum droplets that form in two-component mixtures. We demonstrate that the flattop region emerging in one of the components can serve as a background for the formation of a secondary droplet that is localized on the top of the already existed density plateau. As a result, the total density distribution acquires a double-flattop structure being composed of a compressed internal core surrounded by a rarefied shell. 
The analytical predictions are developed using a perturbation theory, which indicates that in the leading order the shape and stability of the emerging states is determined by the cubic nonlinear Schr\"idinger (NLS) equation. Numerical results are obtained for one- (1D) and two-dimensional (2D) quantum droplets. In the 1D case, the obtained solutions are stable, whereas in the 2D case there is a countable set of families, and only the fundamental one includes  stable solutions.

The content of this paper can be outlined as follows. Section~\ref{sec:model} introduces the general model and presents analytical results regarding the existence and stability of double-flattop states. Section~\ref{sec:num} presents numerical results for 1D and 2D quantum droplets. Section~\ref{sec:concl} concludes the paper.

\section{The model and analytical results} 

\label{sec:model}

We consider a (generically speaking, multidimentional) Bose-Bose mixture governed by the system of coupled equations for macroscopic wavefunctions: 
\begin{equation}
\label{eq:main}
i\frac{\partial \psi_{1,2}}{\partial t} = -\frac{1}{2} \Delta   \psi_{1,2}  +  \frac{\partial E(n_1, n_2)}{\partial n_{1,2}} \psi_{1,2}.
\end{equation}
In these equations  $\Delta$ is the Laplace operator acting in spatial coordinates, $t$ is the evolution time, the number of spatial coordinates and the specific functional form of the energy density   $E(n_1, n_2)$ are determined by the dimensionality of the problem \cite{Petrov2015, Petrov2016}; $n_{1,2} = |\psi_{1,2}|^2$ are local atomic densities. We search for stationary states with different chemical potentials of two components $\mu_{1,2}$ and with different spatial profiles. These solutions have the form  $\psi_{1,2} = e^{-i\mu_{1,2} t} u_{1,2}$, where the functions $u_{1,2}$ are real-valued, time-independent, and hence satisfy the following system of equations:
\begin{equation}
\label{eq:main:stat}
\mu_{1,2} u_{1,2}  = -\frac{1}{2} \Delta u_{1,2}  +  \rev{ \left.\frac{\partial E(n_1, n_2)}{\partial n_{1,2}}\right|_{n_1 = u_1^2, \, n_2 = u_2^2}} u_{1,2}.
\end{equation}

\subsection{Asymptotic expansions}

We start our analysis from the limit where the density is identically zero in one component (say,  $n_1\equiv 0$), and the density of the second component is uniform and corresponds to the maximal possible value which can be achieved asymptotically for the flattop states: $n_2 \equiv \tn_2 = \mathrm{const}>0$.  The specific value of this density can be determined from the condition
\begin{equation}
\label{eq:1}
\frac{\partial E(0, \tn_2)}{\partial n_2}  \tn_2 =  E(0,  \tn_2),
\end{equation} 
and the chemical potential of the second component is fixed as 
\begin{equation} 
\label{eq:tmu2}
\tmu_2 :=  {\partial E(0, \tn_2)} / {\partial n_2}.
\end{equation}

\rev{We notice that the single-component limit introduced above is not always well-defined, because Eq.~(\ref{eq:1}) may or may not have a nontrivial solution depending on the specific functional form of the energy density $E(n_1, n_2)$. In particular, this limit does not exist for 3D quantum droplets introduced in \cite{Petrov2015}, where the presence  of a binary mixture is indispensable for the formation of the droplet in free space. In the meantime, the situation may be different for low-dimensional geometries, where the liquid phase is more ubiquitous \cite{Petrov2016}, and Eq.~(\ref{eq:1}) may indeed have a solution. Therefore, in what follows we focus on one- and two-dimensional quantum droplets and assume that the introduced limit exists.}

Departing from this  limit, we construct solutions with $n_1 \ll 1$ and $|n_2 - \tn_2| \ll 1$ using   asymptotic expansions with respect to a certain small parameter $\vep \ll 1$. Standard perturbation analysis indicates that such solutions can be constructed  as  
\begin{eqnarray}
\label{eq:u1}
u_1 &=& \vep p_1 + \vep^3 p_3 + \ldots,\\[1mm] 
\label{eq:u2}
u_2 &=& {\tn_2}^{1/2}(1+    \vep^2 q_2 + \vep^4 q_4+ \ldots ) .
\end{eqnarray}
Here the small parameter $\vep$ characterizes the deviation of the chemical potential of the first component from its limit:
\begin{equation}
\label{eq:eps}
\vep^2 = \tmu_1  - \mu_1,
\end{equation}
where  $\tmu_1$ will be determined below [see Eq.~(\ref{eq:tmu1})]. The chemical potential of the second component remains fixed due to the uniform density at infinity: hence $\mu_2 = \tmu_2$. Unknown functions $p_j=p_j(X,Y)$ and $q_j = q_j(X,Y)$ depend on  the slow-scale spatial variables, $X = \vep x$, $Y = \vep y$,   and satisfy zero boundary conditions as $X^2 + Y^2 \to \infty$.

Expansions for stationary densities are obtained as $n_{1,2} = u_{1,2}^2$. The perturbation analysis proceeds by using Taylor expansions for the energy density around $n_1 = 0$ and $n_2 = \tn_2$. Introducing, for compactness, the following notation:
\begin{equation}
\label{eq:pd}
\cE_{jk} = \frac{\partial^{j+k} E(0, \tn_2)}{\partial n_1^j \partial n_2^k},
\end{equation}
we write down the first terms of the expansions as follows:
\begin{equation}
\label{eq:taylor}
\begin{array}{lcr}
\displaystyle
\frac{\partial E(n_1, n_2)}{\partial n_1} &=& \cE_{10} + \cE_{20}n_1 + \cE_{11}(n_2 - \tn_2) + \ldots,\\[3mm]
\displaystyle
\frac{\partial E(n_1, n_2)}{\partial n_2} &=& \cE_{01} + \cE_{11}n_1 + \cE_{02}(n_2 - \tn_2) + \ldots
\end{array}
\end{equation}
We substitute  these expansions in stationary equations (\ref{eq:main:stat}) and collect terms emerging at equal powers of $\vep$. The $O(\vep)$ order of equation for $u_1$ fixes the unperturbed value of the chemical potential of the first component:
\begin{equation}
\label{eq:tmu1}
\tmu_1  =  {\partial E(0, \tn_2)} / {\partial n_1} = \cE_{10}.
\end{equation}
The next pair of nontrivial equations reads
\begin{eqnarray}
p_1 &=& \frac{1}{2} \Delta_\mathbf{R}  p_1   - \cE_{20} p_1^3  + 2\cE_{11} \tn_2  p_1q_2,\\[1mm]
\label{eq:q2}
q_2 &=& -\cE_{11} p_1^2 / (2{\tn_2} \cE_{02}),
\end{eqnarray}
where we use $\Delta_\mathbf{R}$ for  the Laplace operator acting in the slow-scale variables $\mathbf{R} = (X, Y)$. The obtained pair of equations combine in a single one which has the form of a time-independent NLS  equation with \emph{purely cubic} nonlinearity:
\begin{equation}
\label{eq:cubic}
\frac{1}{2} \Delta_\mathbf{R} p_1 - p_1 + \sigma p_1^3 = 0, \quad  \mbox{where\quad} \sigma = -\frac{\cH}{\cE_{02}}, 
\end{equation}
and $\cH = \cE_{20} \cE_{02} - \cE_{11}^2$ is the Hessian determinant.

Equation (\ref{eq:cubic}) has radially symmetric solitary-wave solutions if $\sigma>0$. In the 1D case the solution of interest corresponds to the bright soliton $p_1(X) = (2/  \sigma)^{1/2} \sech (2^{1/2} X)$. In the 2D case, the radially symmetric and positive solution  $p_1(X,Y)$ corresponds to the well-known Townes soliton \cite{Townes}, whose analytical expression is unavailable. In addition, in the 2D case there also exists a countable set of ``excited states'' corresponding to solitary waves with finite number of radial zeros (see e.g. \cite{Fibich,Yang}). Below it will be demonstrated that stable solutions are possible only for $\cE_{02} >0$ (because the uniform density in the second component is modulationally  unstable otherwise). Hence   solitary-wave solutions in the first component can be stable  only  if  $\cH <0$.

As follows from Eq.~(\ref{eq:q2}), the sign of $q_2$ is determined by the second-order mixed derivative $\cE_{11}$. According to expansion (\ref{eq:u2}), if $\cE_{11}$ is negative (which is the case considered below), then a denser spot is formed on top of the uniform background of the second species. On the contrary, positive $\cE_{11}$ would correspond to a localized density reduction in the second species, i.e., the two components would tend to separate. We note that these conclusions are consistent with the analogous conditions for the instability against formation of denser states  in the meanfield theory of Bose-Bose mixtures \cite[Ch.~12]{Pethick}.


The perturbation analysis can be continued to next orders of the asymptotic expansions. This requires taking into account third- and higher-order partial derivatives in Taylor expansions (\ref{eq:taylor}). The next pair of nontrivial equations reads  
\begin{widetext} 
	\begin{eqnarray}
	\label{eq:p3}
	\frac{1}{2}\Delta_\mathbf{R} p_3 - p_3  + 3\sigma p_1^2 p_3 = -\frac{\cE_{11}^2p_1^3}{\cE_{02}^2\tn_2}\left( \frac{\cH(D+2)p_1^2}{2\cE_{02}} + D+1\right) +  \left( {\cE_{30}}  - \frac{3\cE_{11}\cE_{21}}{ \cE_{02}} + \frac{3\cE_{11}^2\cE_{12}}{ \cE_{02}^2} - \frac{\cE_{03}\cE_{11}^3}{ \cE_{02}^3}\right)p_1^5/2,\\[2mm]
	\label{eq:q4}
q_4 = \frac{1}{4\cE_{02}\tn_2}\left[  \frac{\sigma (D+2)p_1^2 - \cE_{11} p_1^2/2  -2(D+1)}{\tn_2  \cE_{02}} \cE_{11} p_1^2   + \left(\frac{2\cE_{11}\cE_{12}}{\cE_{02}} - \frac{\cE_{11}^2 \cE_{03}}{\cE_{02}^2} - \cE_{21}\right)  p_1^4 - 4\cE_{11}p_1 p_3 \right].
	\end{eqnarray}
\end{widetext} 
where $D\in \{1,2\}$ is the dimensionality (i.e., the number of spatial variables), and $\cE_{30}, \cE_{21}, \cE_{12}, \cE_{03}$ are third partial derivatives as per notation (\ref{eq:pd}). Equation (\ref{eq:p3}) is linear and inhomogeneous; its left-hand side corresponds to the linearization of the cubic NLS equation [compare Eqs.~(\ref{eq:p3}) and (\ref{eq:cubic})]. The right-hand side  of Eq.~(\ref{eq:p3}) satisfies the solvability condition, and therefore  Eq.~(\ref{eq:p3}) has a unique  localized solution. If $p_3$ is found, then one can compute $q_4$ from Eq.~(\ref{eq:q4}).

A  simple solution for Eqs.~(\ref{eq:p3})--(\ref{eq:q4}) is available for the 1D case (i.e., $D=1$): $p_3(X) = (2/\sigma)^{1/2} [\cC_1 \sech (2^{1/2} X) + \cC_3 \sech^3 (2^{1/2} X)]$, where
\begin{eqnarray}
\cC_1 = \frac{2}{3} \frac{\cE_{02}^2(\cE_{02}\cE_{30} - 3\cE_{11}\cE_{21})  - \cE_{11}^2(\cE_{03}\cE_{11} - 3\cE_{02}\cE_{12})  }{\cE_{02}\cH^2} \nonumber \\
- {\cE_{11}^2} / {( \tn_{2} \cE_{02}\cH)},\qquad\\[3mm]
\cC_3 = - \cC_1/2 + {\cE_{11}^2}/{(2\tn_2  \cE_{02}\cH)}.\qquad
\end{eqnarray}

\subsection{Stability}

To address the linear stability of the   stationary  solutions, we consider perturbed wavefunctions $\psi_{1,2} = [u_{1,2} + \phi_{1,2}]e^{-i\mu_{1,2}t}$, where $\phi_{1,2}$ are small  time-dependent perturbations. The standard linearization procedure transforms the system (\ref{eq:main}) into the following equations for dynamics of the perturbations:
\begin{equation}
\label{eq:linearizaition}
\begin{array}{c}
\displaystyle
i\frac{\partial \phi_{1,2}}{\partial t} = \left[-\frac{1}{2} \Delta    - \mu_{1,2}    +  \frac{\partial E }{\partial n_{1,2}} \right] \phi_{1,2} \\[6mm] \displaystyle + 2u_{1,2}^2 \frac{\partial^2 E}{\partial n_{1,2}^2}\RE\, \phi_{1,2} + 2u_1 u_2 \frac{\partial^2 E }{\partial n_{1}\partial n_2} \RE\, \phi_{2,1},
\end{array}
\end{equation}
where the partial derivatives of the function $E(n_1, n_2)$ are evaluated at stationary densities $n_{1,2} = u_{1,2}^2$.

First, we check the modulational instability of the uniform solution $u_1\equiv 0$, $u_2 \equiv  \tn_2^{1/2}$ which corresponds to $\vep=0$ in our asymptotic expansions. In this limit the equations (\ref{eq:linearizaition}) decouple. The equation for $\phi_1$ corresponds to trivial and stable dynamics, while the second perturbation evolves as 
\begin{equation}
i\frac{\partial \phi_{2}}{\partial t} = -\frac{1}{2} \Delta \phi_{2}   + 2\tn_2\cE_{02}  \RE\, \phi_{2}.
\end{equation}
The obtained equation is well-known in theory of nonlinear waves, as it describes the dynamics of   the modulational instability of constant-amplitude solutions in the cubic NLS equation \cite{Yang}, where the coefficient $\cE_{02}$ determines the sign of the effective nonlinearity. For the NLS equation it is well-known that the constant-intensity solutions are stable for  $\cE_{02} > 0$, which corresponds to the repulsive nonlinearity in the meanfield theory of BECs. Hence condition $\cE_{02} > 0$ is necessary for the solutions to be stable.  

Second, we consider the linear stability equations (\ref{eq:linearizaition}) in the limit of small $\vep$, where solutions $u_{1,2}$ obey the asymptotic expansions developed above. Different orders of leading corrections in expansions (\ref{eq:u1}) and (\ref{eq:u2}) suggest that the characteristic amplitudes of perturbations are also different for the first and second species. We therefore assume the following substitutions:  $\phi_1 = \Phi_1(T, X, Y)$ and $\phi_2 = \vep \Phi_2(T, X, Y)$, where $X,Y$ are slow spatial variables introduced above and $T = \vep^2 t$ is the slow time. Then, in the leading order of the perturbation theory we obtain the following equation:
\begin{equation}
i\frac{\partial \Phi_{1}}{\partial T} = -\frac{1}{2} \Delta_\mathbf{R} \Phi_1 + \Phi_1 - \sigma p_1^2(\Phi_1 + 2\RE\,\Phi_1).
\end{equation}
This is the familiar equation that determines the linear stability of a solitary wave $p_1(X,Y)e^{-iT}$, provided that its dynamics is governed by the time-dependent cubic NLS equation, see e.g.  \cite[Ch.~5.2]{Yang} or \cite[Ch.~9.2]{Fibich}. Thus, in the leading order of the perturbation theory,  the linear stability of   solutions obeying asymptotic expansions (\ref{eq:u1})-(\ref{eq:u2}) is determined by the rather well-studied stability of solitary waves in the NLS equation. In particular, we immediately conclude that the solutions are expected to be stable in the 1D case, where  robustness of the  bright soliton $p_1(X)$ is guaranteed by the complete integrability of the NLS equation. 

The 2D case is more delicate, because  the Townes soliton (i.e., the positive and radially symmetric solution $p_1(X,Y) >0$) is stable in the framework of the linear stability analysis, but is nonlinearly unstable in the time-dependent NLS equation, where the finite-time collapse can be observed in evolution of  slightly perturbed Townes states \cite[Ch.~5.10]{Yang}. In the quantum-droplet system, the finite-time collapse is suppressed, and the nonlinear dynamics is expected to be significantly different. Indeed, the eventual finite-time collapse implies the extreme narrowing of the wavefunction and hence drives the solution away from the regime where the asymptotic expansions (\ref{eq:u1})-(\ref{eq:u2}) are valid, because the latter are obtained under the assumption of  smoothly changing wavefunctions. Regarding the ``excited'' radially symmetric 2D states with finite number of radial zeros, linearization of the 2D NLS equation predicts that these solutions are linearly unstable with respect to azimuthal perturbations. Hence the corresponding quantum droplets are expected to be unstable too.

\section{Numerical results} 
\label{sec:num}

Asymptotic expansions constructed above predict the existence of spinor states composed of a localized wave in the first component coupled to a localized dense spot situated on the background of infinite extent in the second component. For these solutions the chemical potential $\tmu_2$ of the second species is fixed by the  density background, as per Eq.~(\ref{eq:tmu2}). However, tuning the chemical potential of the second component away from the limiting value $\tmu_2$, one can transform the spatially infinite background into a finite-size flattop droplet. Therefore in what follows, we present numerical results for spinor droplets that obey zero boundary conditions in \textit{ both components}: $\lim_{x^2+y^2  \to \infty} u_{1,2} = 0$. The results will be presented for one- and two-dimensional droplets. Asymptotic expansions developed above can be used to prepare a  suitable initial guess  for Newton iterations to converge to a localized solution. Using the numerical continuation, we depart from the   limit $\vep\to 0$ and address also the regime, where the peak density of the first component is no longer small.

\subsection{One-dimensional solutions}

For quasi-one-dimensional quantum droplets, the energy density has the form   \cite{Petrov2016, Kartashov2022,Mithun2020}:
\begin{eqnarray}
E(n_1, n_2) = \frac{(g_{1}^{1/2} n_1   - g_{2}^{1/2} n_2)^2}{2} - \frac{2}{3\pi} (g_{1} n_1 + g_{2} n_2)^{3/2} \nonumber\\[1mm]
+ \frac{\delta (g_{1} g_{2})^{1/2}}{(g_{1} + g_{2})^2} (g_{1}^{1/2} n_2  + g_{2}^{1/2}n_1)^2 .\qquad
\label{eq:E1D}
\end{eqnarray}
Here dimensionless coefficients $g_{1,2}>0$ characterize intra-species interactions in each component, the coefficient $g_{12}<0$ defines inter-species interactions and we introduce auxiliary constant $\delta = g_{12} +  ( g_{1} g_{2})^{1/2}$. In the experimental situation, $\delta$ is positive and small. In a generic case $g_1$ and $g_2$ are different. For our numerical results we adopt the following  normalized values which correspond to a mixture of two species of $^{39}$K atoms in different hyperfine states \cite{Kartashov2022}:  $g_{1} =  0.639$, $g_{2} = 2.269$ and $g_{12} = -1$.

Solving Eq.~(\ref{eq:1}) we find that the maximal density corresponding to the flattop solution in the second component is equal to $\tn_2 \approx 9.891 \times 10^{-2}$. The limiting chemical potentials can be computed as 
\begin{equation}
\tmu_1 \approx -0.209, \quad  \tmu_2 \approx -0.114.
\end{equation}
Computing the second-order derivatives of the energy density and the Hessian determinant at $n_1=0$ and $n_2 = \tn_2$ , we check that
\begin{equation}
\label{eq:ineq}
\cE_{02}>0, \quad \cE_{20}>0, \quad \cE_{11}<0, \quad \cH<0,
\end{equation}
i.e., the conditions for the asymptotic expansions to apply hold.

\begin{figure}
	\begin{center}
		\includegraphics[width=0.999\columnwidth]{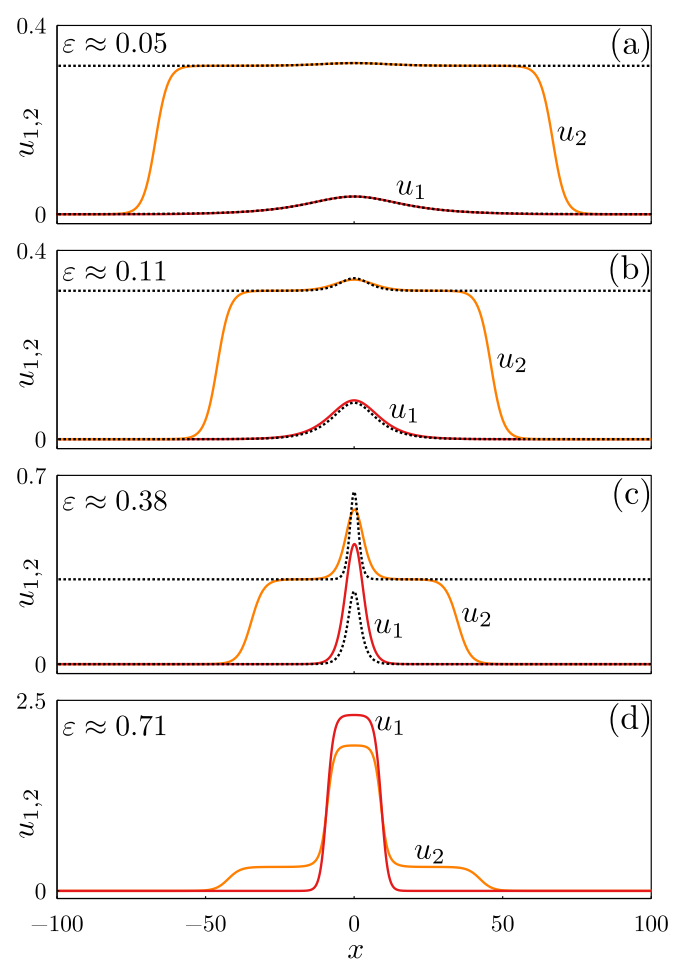}
	\end{center}
	\caption{Profiles of  1D quantum droplets emerging from the uniform solution $u_1\equiv 0$, $u_2 \equiv \tn_2$ when the chemical potential of the second species is slightly below the limiting value $\tmu_2$; specifically $\tmu_2 - \mu_2 \approx 10^{-7}$ in this figure. Four panels correspond to the increase of the formal parameter $\vep$ and, respectively, to the decrease of  the chemical potential of the first component departing from its limiting value $\tmu_1$, see Eq.~(\ref{eq:eps}) for definition of $\vep$. The specific values of $\vep$ are indicated in the panels.  Red and orange curves show   first ($u_1$) and second ($u_2$) components, and  doted lines in (a,b,c) show the solutions obtained from the truncated asymptotic expansions at the corresponding values of  $\vep$.}
	\label{fig:1d}
\end{figure}

Solutions satisfying zero boundary conditions for both atomic species are found to exist for chemical potentials $\mu_2$ slightly below the limiting value $\tmu_2$. The examples of these solutions are presented in Fig.~\ref{fig:1d} for the fixed  value of $\mu_2<\tmu_2$ and for several increasing values of the parameter $\vep$ (which corresponds to decreasing chemical potential $\mu_1$ of the first species, see Eq. (\ref{eq:eps})).   

For small $\vep$  in Fig.~\ref{fig:1d}(a-c), we compare the numerical localized solutions with the shapes obtained from the perturbation analysis. The initial increase of $\vep$ (resp., the decrease of the   chemical potential $\mu_1$) leads to the growth of the number of particles in the first species, while the  flattop region in the second component becomes slightly narrower, cf. Fig.~\ref{fig:1d}(a) and (c). At the same time,   the decrease of the chemical potential $\mu_1$ results in the formation of a secondary dense spot in both components, so that eventually the density profile of the second component and the total density $n_1 + n_2$ acquire a \textit{double-flattop} form featuring two plateaus of different heights, as displayed in Fig.~\ref{fig:1d}(d). The formation of double-flattop solutions   limits the interval of chemical potentials $\mu_1$, where the solutions of this type can be found: no further decrease of $\mu_1$ is possible as the number of particles in each component diverges to infinity.

As the chemical potential $\mu_2$ is decreased further from the limiting value $\tmu_2$, the outer flattop region gradually shrinks, and eventually the solutions lose the distinctive double-flattop shape. For sufficiently small $\mu_2$ each component becomes bell-shaped and resembles  the conventional bright soliton. Since two-component quantum droplets of this form  have been studied in earlier literature \cite{Mithun2020,Mista2021}, we do not consider them here.
 
\begin{figure}
	\begin{center}		\includegraphics[width=0.999\columnwidth]{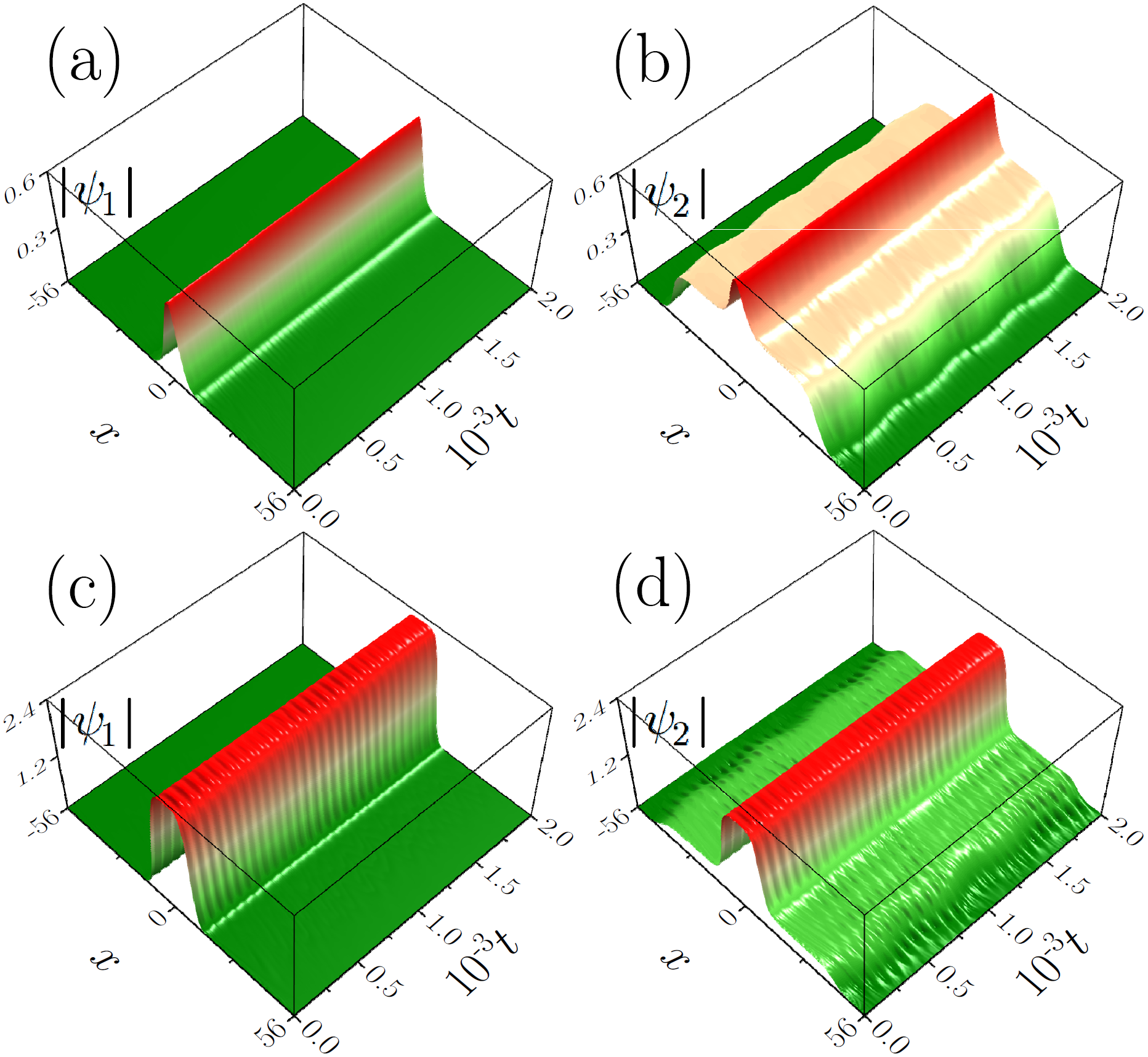}\\%
	\end{center}
	\caption{Nonlinear evolution of 1D quantum droplets from Fig.~\ref{fig:1d}(c) (upper panels) and Fig.~\ref{fig:1d}(d) (lower panels). Left and right panels display the amplitudes of wavefunctions of the first ($|\psi_1|$) and second ($|\psi_2|$) species, respectively.}
	\label{fig:dynam1d}
\end{figure} 

The preliminary stability analysis performed above indicates that the 1D solutions on the infinite background are linearly stable, at least for small values of $\vep$. In order to confirm the stability for self-bound droplets satisfying zero boundary conditions in both components, we solved the linear stability problem numerically. To this end we return to the linear stability equations (\ref{eq:linearizaition}), and use the standard substitution for the perturbations:    $\phi_{1,2}(x,t) = \zeta_{1,2}(x) e^{\lambda t}  +  \eta^*_{1,2}(x) e^{\lambda^* t}$, where the asterisks denote complex conjugation. Then   the linearization equations  (\ref{eq:linearizaition})  transform into an eigenvalue problem for  $\lambda$. The linear stability takes place if the linearization spectra does not contain eigenvalues $\lambda$ with positive real part. The complete set of equation constituting the linear stability eigenproblem can be found in \cite{KartashovZ2024} and we therefore do not duplicate it here.

Numerical evaluation of the linearization spectra confirms the stability of solutions shown in Fig.~\ref{fig:1d}. In addition, we have modelled the nonlinear dynamics governed by the time-dependent equation (\ref{eq:main}) with  the energy density given by Eq.~(\ref{eq:E1D}) for  several initial conditions corresponding to the 1D quantum droplets. To promote the potential development of eventual instabilities, artificial perturbations have been added to numerical solutions $u_{1,2}(x)$ at $t=0$ in the form of a complex-valued random noise whose maximal amplitude is about $5\%$ of the solution amplitude. These simulations also confirm the stability of the obtained solutions. The examples of dynamics near the double-flattop regime and at the double-flattop regime are presented in Fig.~\ref{fig:dynam1d}(a,b) and Fig.~\ref{fig:dynam1d}(c,d), respectively.

\subsection{Two-dimensional solutions}

For 2D quantum droplets, we limit the numerical study by the case of symmetric coupling constants \cite{Li2018, Kartashov2020, Petrov2016}. The corresponding energy density reads
\begin{equation}
\label{eq:E2D}
E(n_1, n_2) = \frac{(n_1 - n_2)^2}{2} +  \frac{\alpha}{2}(n_1+n_2)^2\ln\frac{n_1+n_2}{\sqrt{e}},
\end{equation}
where $\alpha$ is the coupling coefficient. For numerical results presented below we assume the normalization with $\alpha=1$ \cite{Kartashov2020}. Density $\tn_2$ and corresponding chemical potentials $\tmu_{1,2}$ can easily be found for this case and inequalities (\ref{eq:ineq}) can also be checked.

\begin{figure}
	\begin{center}
		\includegraphics[width=0.999\columnwidth]{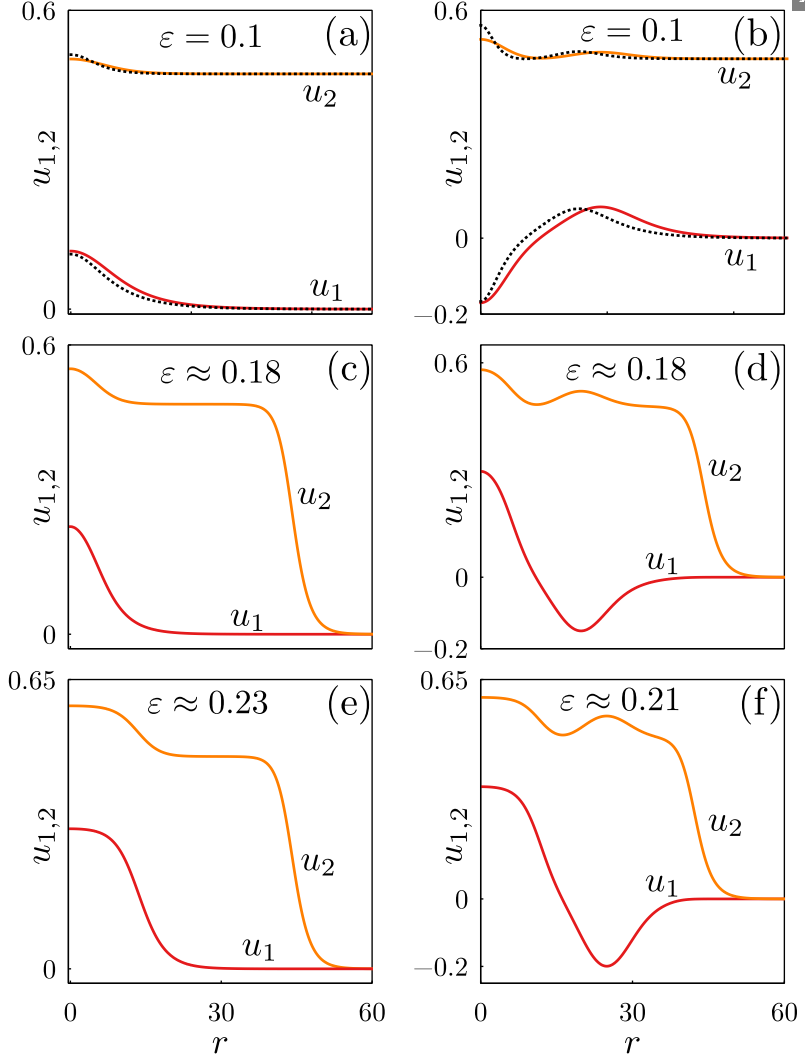}
	\end{center}
	\caption{(a,b) Radial profiles of 2D  quantum droplets emerging from the uniform solution $u_1\equiv 0$, $u_2 \equiv \tn_2$ when the chemical potential of the second species is exactly equal to the limiting value $\tmu_2$. Red and orange curves show first ($u_1$) and second ($u_2$) components obtained numerically, while dotted lines in upper panels show the solutions obtained from the truncated  asymptotic expansions at corresponding value of $\vep$. (b,c) Numerically computed localized quantum droplets for $\mu_2$ above the $\tmu_2$ value (specifically, at $\mu_2 - \tmu_2 \approx 2.295 \times 10^{-3}$). Left and right panels correspond to the ``ground state'' (corresponding to the nodeless Townes mode $u_1$ in the leading order of the perturbation theory) and to the ``excited state'' with one radial zero in $u_1$ component, while $u_2$ component is always nodeless in our case.}
	\label{fig:2d}
\end{figure}

In agreement with predictions of the perturbation theory, in the 2D case we have found several families of radially-symmetric solutions with different numbers of radial zeros in the wavefunction of the first species. In Fig.~\ref{fig:2d}(a) we present the nodeless (``ground state'') solutions and the single-node (``first excited state'') solutions computed for infinitely extended second species and compare them with the shapes obtained from the asymptotic expansions.

When the chemical potential of the second species is different from the limiting value $\tmu_2$, one can obtain localized quantum droplets satisfying zero boundary conditions for wavefunctions of both components: $\lim_{x^2  + y^2 \to \infty}u_{1,2} = 0$. 
Specifically, the 2D solitary droplets can be found   for chemical potentials above the corresponding value, i.e., for $\mu_2 > \tmu_2$. Radial profiles of localized ``ground states'' and  ``excited states'' are shown in Fig.~\ref{fig:2d}(c,e) and  Fig.~\ref{fig:2d}(d,f), respectively. Similarly to the 1D case, for ground states the decrease of the chemical potential $\mu_1$ results in the formation of the double-flattop quantum droplets featuring two distinct regions with nearly uniform density, see Fig.~\ref{fig:2d}(e).

To examine the linear stability of 2D solutions, we use the linearization equations (\ref{eq:linearizaition}) with the perturbations represented as $\phi_{1,2}(x,y,t) = \sum_{m=0}^\infty [ \zeta_{1,2,m}(r) e^{im\varphi   + \lambda t} + \eta^*_{1,2,m}(r) e^{-im \varphi + \lambda^*t}]$, where the functions $\zeta_{1,2,m}$ and  $\eta_{1,2,m}$  depend only on the polar radius $r = (x^2 + y^2)^{1/2}$, $\varphi$ is the polar angle, and integer $m$ stands for the azimuthal index of the perturbation. The equations corresponding to different perturbation indices $m$ should be solved for $\lambda$ and radial profiles $\zeta(r)$ and $\eta(r)$. The complete set of the linearized equations for perturbations can be found in Ref.~\cite{Kartashov2020}. Assuming that eventual instabilities correspond to a limited range of azimuthal perturbation indices, we have solved the linear stability eigenproblem for $m=0,  1, \ldots,   10$. This study confirms linear stability of ground-state quantum droplets, while the states with radial nodes were found to be unstable. For our parameters the largest perturbation growth rates were obtained for azimuthal perturbation indices $m=4$ and $m=5$. 

\begin{figure}
	\begin{center}
		\includegraphics[width=0.999\columnwidth]{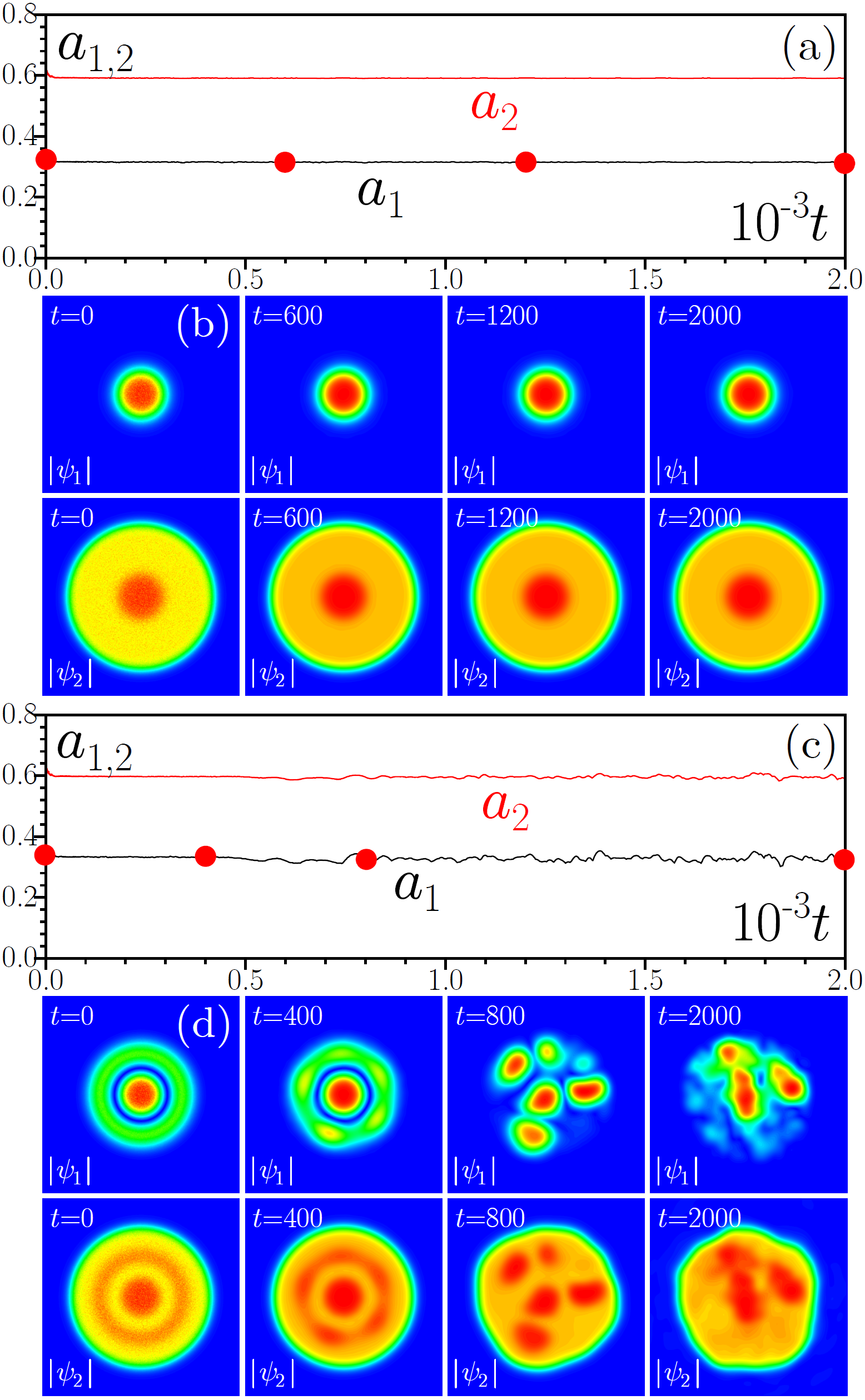}%
	\end{center}
	\caption{Evolution dynamics of the perturbed 2D quantum droplets without radial nodes (a),(b) and with radial node in the first component (c),(d). Maximal amplitudes of two components of the wavefunction versus time are shown in (a),(c), while amplitude distributions of the wavefunction corresponding to the red dots are shown in (b),(d). All amplitude distributions are shown within the window $(x,y)\in [-60, 60]\times[-60, 60]$.}
	\label{fig:dynam2d}
\end{figure}

Numerical integration of the time-dependent equations (\ref{eq:main}) with the energy density given by Eq.~(\ref{eq:E2D}) indicates that the nodeless states are also nonlinearly stable and confirms the dynamical instability of the excited states. Examples of stable and unstable dynamics of the ground and excited states are shown in Fig.~\ref{fig:dynam2d}. Specifically, Figs.~\ref{fig:dynam2d}(a,c) display the evolution of the maximal amplitude of each wavefunctions: $a_{1,2}(t) = \max_{(x,y)} |\psi_{1,2}|$, and Figs.~\ref{fig:dynam2d}(b,d) show several snapshots of the amplitude distribution. For the excited state we observe that the initial onset of the azimuthal instability (snapshot at $t=400$) eventually destroys the radial structure of the first component (snapshots at $t\geq 800$), while the droplet in the second  component remains relatively robust, and the maximal amplitudes $a_{1,2}(t)$ feature irregular oscillations around their initial values.

\section{Discussion and conclusion}
\label{sec:concl}

The formation of flattop shapes, i.e.,  the regions of  almost uniform condensate density, is a  representative feature of quantum droplets. In this paper, we have introduced a class of double-flattop quantum droplets whose total density features two, rather than one, distinctive density plateaus. The existence and stability of solutions of this form are predicted using the asymptotic expansions that depart from the limit, where the density of one species is identically zero. Close to this limit, the problem is reduced to the well-studied nonlinear Schr\"odinger equation which can be used to obtain the prediction about the possible shapes of solutions and on their stability. Using the numerical continuation, the solutions are extrapolated into the regime where the maximal densities of both  species are comparable. The predicted solutions essentially exploit the two-component nature of the Bose-Bose mixture and cannot be found in the reduced case under the assumption of identically shaped density distributions in both species. The numerical results are presented for one- and two-dimensional droplets, and stable double-flattop solutions are found in either case.

Our findings enrich the family of quantum droplets emerging in Bose-Bose mixtures and open several directions for future research. In particular, in the 2D case, the analytical and numerical results can be potentially extended onto solutions bearing nonzero vorticity in one of the components. While the present paper addresses mainly stationary states, the perturbation theory can  be extended on time-dependent regularly breathing solutions. In this case, the leading-order behavior reduces to the time-dependent NLS equation, and in the 1D case one can expect to encounter intriguing  quasi-integrable dynamics for quantum droplets. In addition, a plethora of interesting phenomena that are known to exist for quantum droplets confined by external potentials \cite{Liu2019, Zhou2019, Zhang2019, Dong2020, Morera2020, Zheng2021, Dong2022, Pathak2022, Liu2022, Zezyulin2023, Liu2023, Nie2023, Huang2023, Kartashov2024} suggest the possibility of existence of such states in the presence of harmonic confinement or optical lattices.

\section*{Declaration of competing interest}

The authors declare that they have no known competing financial interests or personal relationships that could have appeared to influence the work reported in this paper.

\section*{Funding}

Y.V.K. acknowledges funding by the research project FFUU-2024-0003 of the Institute of Spectroscopy of the Russian Academy of Sciences. The work of D.A.Z.  was supported by   the Ministry of Science and Higher Education of the Russian
Federation, goszadanie no. 2019-1246, and by    Priority 2030 Federal Academic Leadership Program.

\section*{CRediT authorship contribution statement}
\textbf{Yaroslav V. Kartashov}: Writing – original draft, Visualization, Methodology, Investigation, Conceptualization. \textbf{Dmitry A. Zezyulin}:
Writing – original draft, Visualization, Methodology, Investigation, Conceptualization.

\section*{Data availability}

Data will be made available on request.



\end{document}